\documentclass[a4paper]{spie}  

\usepackage{amsmath,amsfonts,amssymb}
\usepackage{graphicx}
\usepackage[colorlinks=true, allcolors=blue]{hyperref}
\usepackage{lipsum}

\title{Measuring the CMB spectral distortions with COSMO: the multi-mode antenna system}

\author[1,2]{E. Manzan}
\author[1]{L. Albano}
\author[1,2]{C. Franceschet}
\author[3,4]{E. S. Battistelli}
\author[3,4]{P. de Bernardis}
\author[1,2]{M. Bersanelli}
\author[3,4]{F. Cacciotti}
\author[3,4]{A. Capponi}
\author[3,4]{F. Columbro}
\author[5,6]{G. Conenna}
\author[5,6]{G. Coppi}
\author[3,4]{A. Coppolecchia}
\author[3,4]{G. D’Alessandro}
\author[3,4]{G. De Gasperis}
\author[3,4]{M. De Petris}
\author[5,6]{M. Gervasi}
\author[3,4]{G. Isopi}
\author[3,4]{L. Lamagna}
\author[5,6]{A. Limonta}
\author[3,4]{E. Marchitelli}
\author[3,4]{S. Masi}
\author[1,2]{A. Mennella}
\author[1]{F. Montonati}
\author[5,6]{F. Nati}
\author[3,4]{A. Occhiuzzi}
\author[3,4]{A. Paiella}
\author[7]{G. Pettinari}
\author[3,4]{F. Piacentini}
\author[8]{L. Piccirillo}
\author[3,4]{G. Pisano}
\author[9]{C. Tucker}
\author[5,6]{M. Zannoni}

\affil[1]{Università degli studi di Milano, Milano, Italy}
\affil[2]{INFN sezione di Milano, 20133 Milano, Italy} 
\affil[3]{Università di Roma - La Sapienza, Roma, Italy}
\affil[4]{INFN sezione di Roma, 00185 Roma, Italy} 
\affil[5]{Università di Milano - Bicocca, Milano, Italy} 
\affil[6]{INFN sezione di Milano - Bicocca, 20126 Milano, Italy} 
\affil[7]{Istituto di fotonica e nanotecnologie, Consiglio Nazionale delle Ricerche IFN-CNR, 00133 Roma, Italy} 
\affil[8]{Jodrell Bank Centre for Astrophysics, School of Physics and Astronomy, University of Manchester, Manchester, UK} 
\affil[9]{School of Physics and Astronomy, Cardiff University, Cardiff, UK} 

\authorinfo{Further author information: \\ Elenia Manzan: E-mail: elenia.manzan@unimi.it}

\begin{document} 
\maketitle

\begin{abstract}
In this work, we present the design and manufacturing of the two multi-mode antenna arrays of the COSMO experiment and the preliminary beam pattern measurements of their fundamental mode compared with simulations.

COSMO is a cryogenic Martin-Puplett Fourier Transform Spectrometer that aims at measuring the isotropic y-type spectral distortion of the Cosmic Microwave Background from Antarctica, by performing differential measurements between the sky and an internal, cryogenic reference blackbody. To reduce the atmospheric contribution, a spinning wedge mirror performs fast sky-dips at varying elevations while fast, low-noise Kinetic Inductance detectors scan the interferogram.

Two arrays of antennas couple the radiation to the detectors. Each array consists of nine smooth-walled multi-mode feed-horns, operating in the $120-180$ GHz and $210-300$ GHz range, respectively. The multi-mode propagation helps increase the instrumental sensitivity without employing large focal planes with hundreds of detectors. The two arrays have a step-linear and a linear profile, respectively, and are obtained by superimposing aluminum plates made with CNC milling. The simulated multi-mode beam pattern has a $\sim 20^{\circ} - 26^{\circ}$ FWHM for the low-frequency array and $\sim 16^{\circ}$ FWHM for the high-frequency one. The side lobes are below $-15$ dB.


To characterize the antenna response, we measured the beam pattern of the fundamental mode using a Vector Network Analyzer, in far-field conditions inside an anechoic chamber at room temperature. We completed the measurements of the low-frequency array and found a good agreement with the simulations. We also identified a few non-idealities that we attribute to the measuring setup and will further investigate. A comprehensive multi-mode measurement will be feasible at cryogenic temperature once the full receiver is integrated.

\end{abstract}




\section{INTRODUCTION}
\label{sec:intro}  

The Cosmic Microwave Background (CMB) is a relic radiation from the early stage of the Universe, when radiation was initially coupled with primordial plasma. About 380000 years after the Big Bang, as the temperature fell below $T\sim10^3\ $K, matter became neutral, and radiation decoupled, propagating almost freely in the expanding universe. The CMB is currently detected as a blackbody spectrum with a brightness peak around 160 GHz \cite{Mater1990} and has proven to be a powerful tool to constrain the cosmological parameters to sub-percent precision \cite{Planck2020}.

However, small deviations from a pure blackbody spectrum are expected, resulting from perturbations in the thermodynamic equilibrium, caused by energy injections during the thermal history of the Universe. Several mechanisms of energy injections are theoretically predicted, both within the $\Lambda$CDM model and by more exotic scenarios \cite{Chluba_2011}. CMB spectral distortions are thus a powerful tool for investigating the thermal history of the Universe and testing theories. 

The brightest distortion predicted by standard cosmology is due to the inverse Compton scattering of CMB photons off free electrons during the Reionization, namely, the epoch when the first stars formed and the matter was ionized once more, and in regions with structure formation. The expected distortion is isotropic and its amplitude is parameterized by the dimensionless Compton factor, $y$, which according to the theory is $y\sim 10^{-6}$ \cite{Hill_2015}. A non-detection of this signal would directly point toward the need for new physics.

The current best upper limit on $y$ is $y < 1.5 \cdot 10^{-5}$, set by COBE-FIRAS \cite{Fixsen_1996} and TRIS \cite{Gervasi_2008}. Detecting this signal requires instruments with high sensitivity, excellent control over instrumental systematic effects, and spurious signals, such as atmospheric emission and astrophysical foregrounds. In this context, COSMO \cite{Masi_2021_cosmo} represents the next generation of ground-based experiments for measuring the CMB spectral distortions. It aims at reaching a sensitivity to the isotropic y-distortion of $\sim10^{-6}$ \cite{Mele_2022}, thus improving the state-of-the-art by an order of magnitude and potentially performing the first detection of the reionization distortion.

This paper is organized as follows: in Section \ref{sec:COSMO} we provide a brief overview of the COSMO experiment; in Section \ref{sec:Antennas} we present the design and simulated performance of the COSMO multi-mode antenna system; in Section \ref{sec:measurement} we describe the measurement strategy used to perform a first characterization of the antenna response and the preliminary results. We summarize our findings and future developments in Section \ref{sec:conclusions}.

\section{The COSMO experiment}
\label{sec:COSMO}  

COSMO is a cryogenic Martin-Puplett Fourier Transform Spectrometer that will operate from Dome-C, in Antarctica. It will measure the difference in brightness between the radiation collected from the sky and an internal, cryogenic reference blackbody.

\begin{figure} [ht]
   \begin{center}
   \begin{tabular}{c c} 
   \includegraphics[height=5cm]{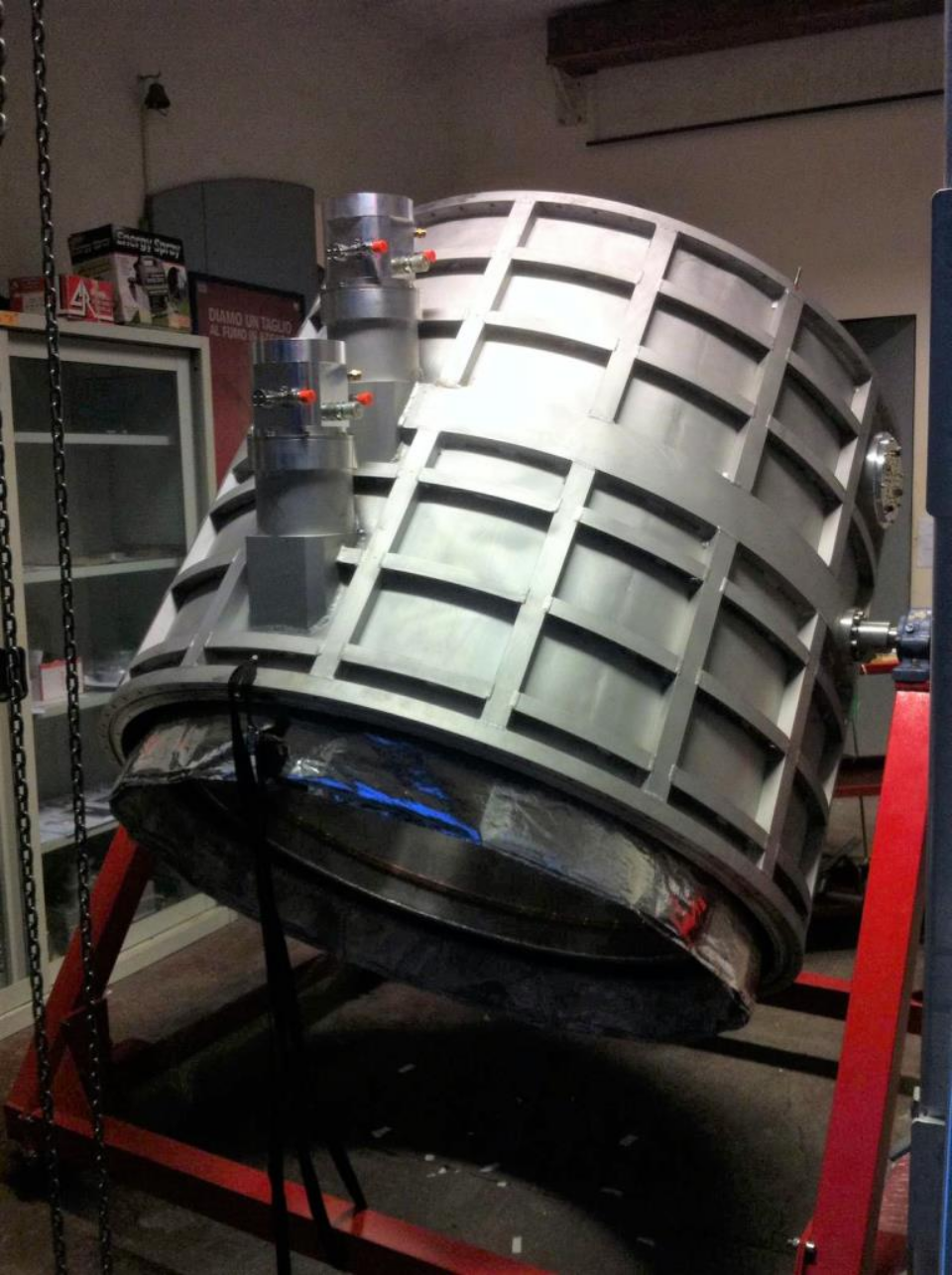}
    \includegraphics[height=5cm]{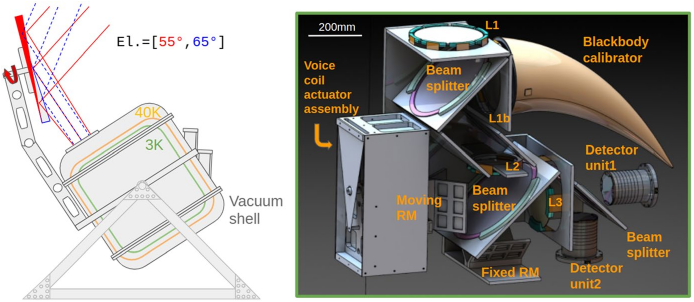}
   \end{tabular}
   \end{center}
   \caption[example] 
   {\label{fig:cosmo}\textit{Left panel:} picture of the COSMO cryostat placed on a mount that allows sweeping in elevation. \textit{Central panel:} schematic of the overall system made by the cryostat, the elevation mount, and a spinning wedge mirror that allows scanning circles in the sky. \textit{Right panel:} CAD design of the instrument inside the cryostat. The first lens, L1, transmits the sky signal inside the cryostat, while a twin lens, L1b, transmits the signal from the internal calibrator. Two beam splitters separated by a second lens, L2, split the radiation onto the interferometer, where two roof mirrors (RM) reflect it toward a third lens, L3, and a last beam splitter, which focuses the radiation onto the two focal planes. A voice coil allows the movement of one of the two RMs to introduce an optical path difference.}
   \end{figure}

The COSMO cryostat and the instrument design are shown in Fig.~\ref{fig:cosmo}. A system of lenses (L1, L1b, L2, L3), beam splitters, and two roof mirrors (RM) combine the radiation onto two focal planes, one at $150$ GHz and the other at $250$ GHz. One of the roof mirrors can be moved through a voice coil to introduce an optical path difference between the interferometer arms. This creates an interference pattern on the focal planes from which the CMB spectrum is recovered with a $\sim5$ GHz spectral resolution.

The altitude and cold, dry climate of the Antarctic plateau make it one of the best sites in the world to perform CMB observations. Although the water vapor content in the atmosphere is low, its emission in the microwave range is not negligible. To remove the atmospheric contribution and reduce the noise due to its fluctuations, COSMO performs fast sky-dips at varying elevations through a spinning wedge mirror, during which the interferogram is scanned by fast, low-noise Kinetic Inductance Detectors (KID).

The radiation is coupled to the focal planes by two arrays of feed-horns, each followed by a flange of linear flares that directly illuminate the KIDs. The antennas, the flares, and the detectors are multi-mode. Since they receive more power than a traditional single-mode receiver, a higher signal-to-noise ratio per detector can be achieved, thus reaching a high instrumental sensitivity with a small number of detectors.

\section{The COSMO multi-mode antenna system}
\label{sec:Antennas}  

\begin{figure} [ht]
   \begin{center}
   \begin{tabular}{c c c} 
   \includegraphics[height=4cm]{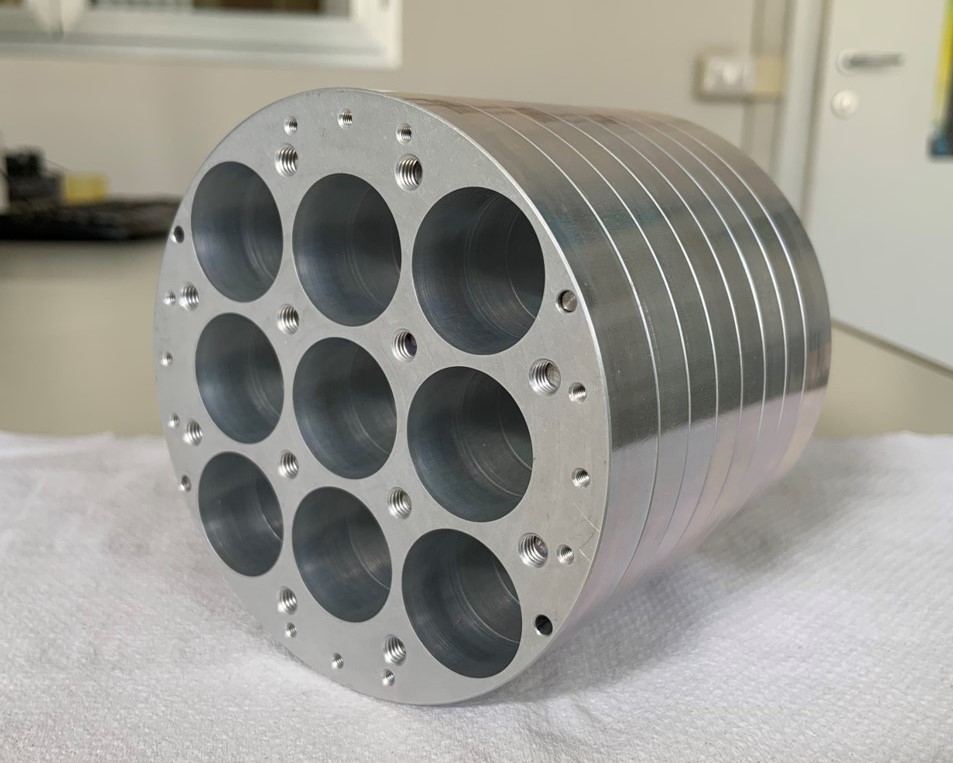}
    \includegraphics[height=4cm]{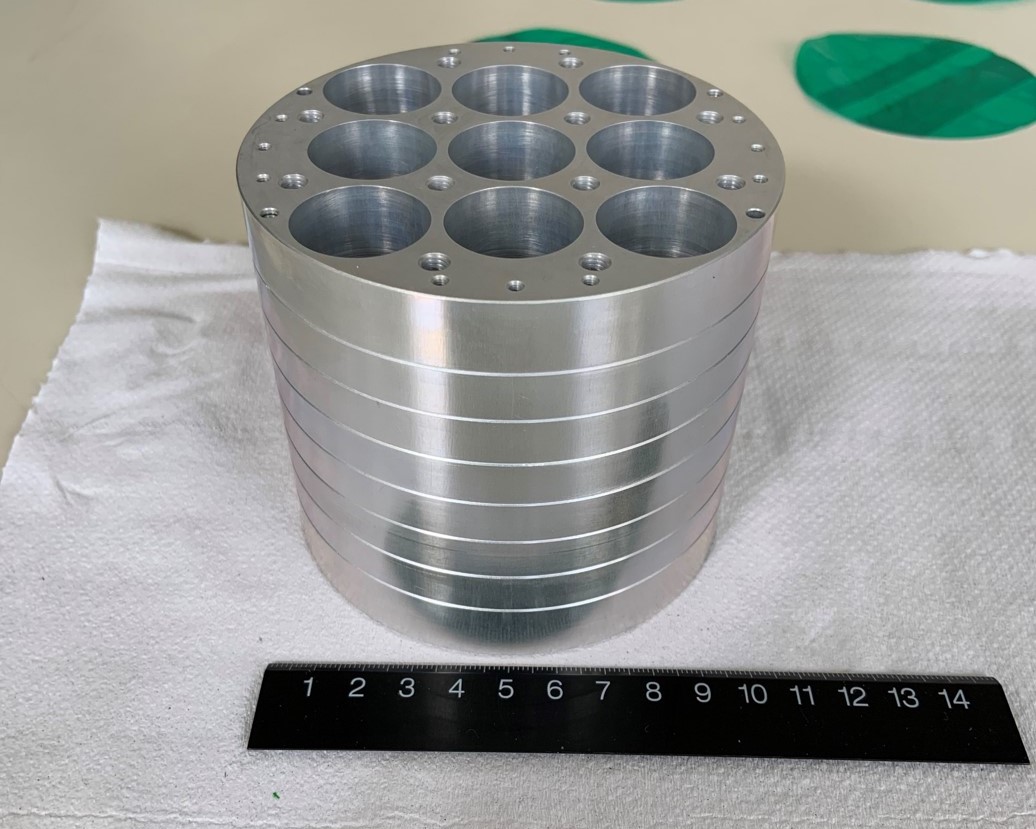}
    \includegraphics[height=4cm]{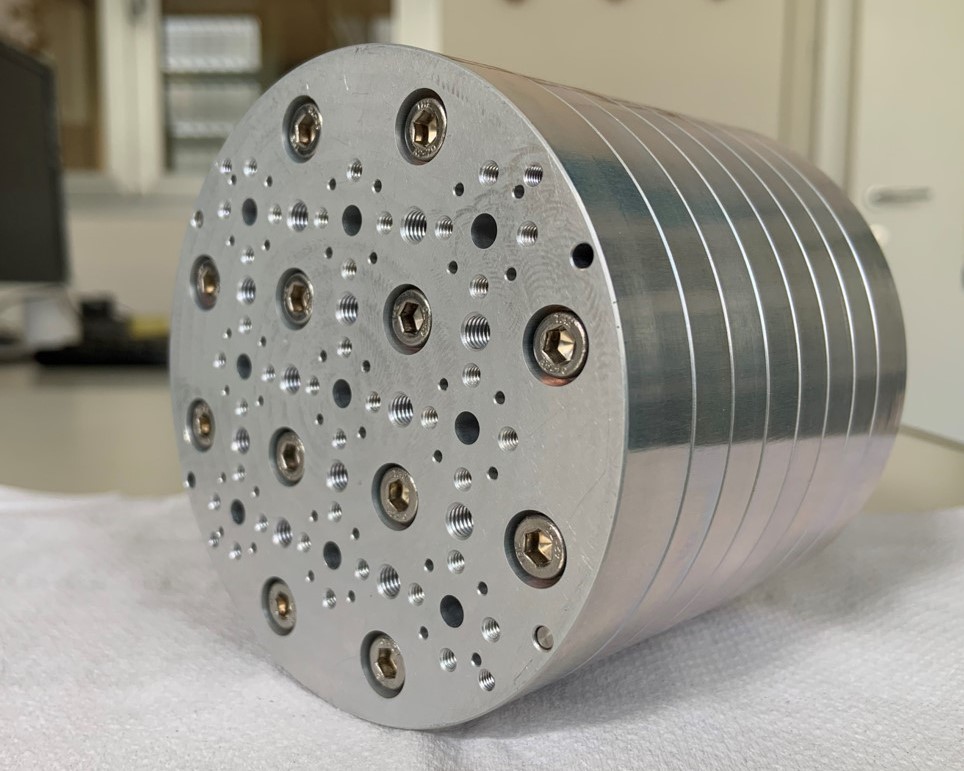}
   \end{tabular}
   \end{center}
   \caption[example] 
   {\label{fig:array1_pictures}Front, side, and back view of the COSMO 150 GHz array.}
   \end{figure}

\begin{figure} [ht]
   \begin{center}
   \begin{tabular}{c c c} 
   \includegraphics[height=4cm]{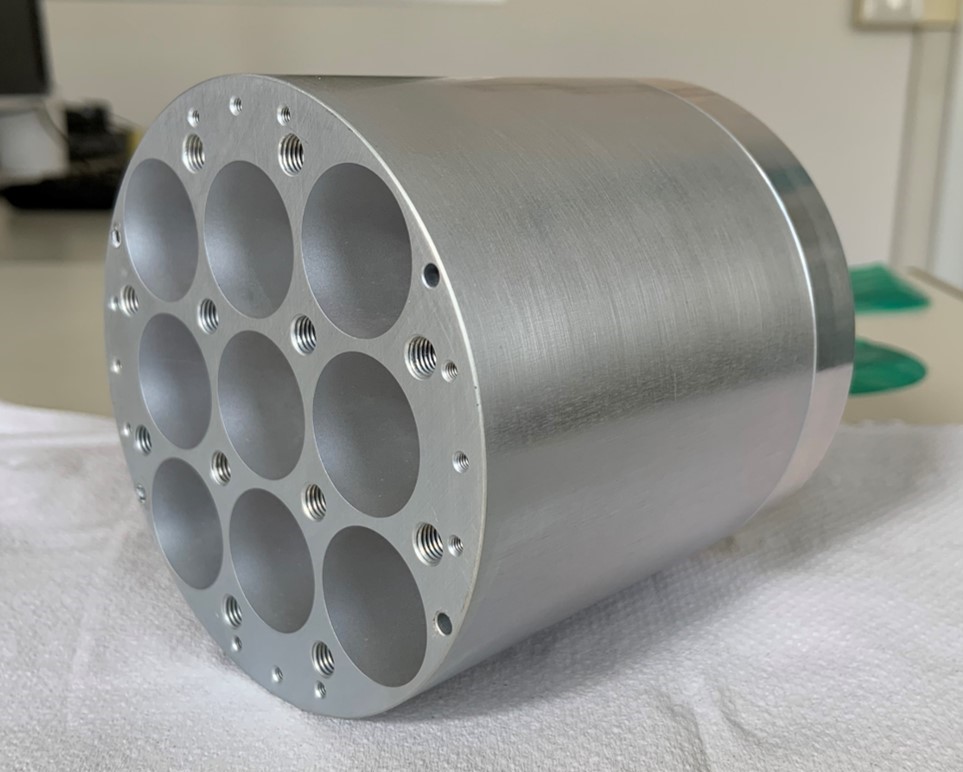}
    \includegraphics[height=4cm]{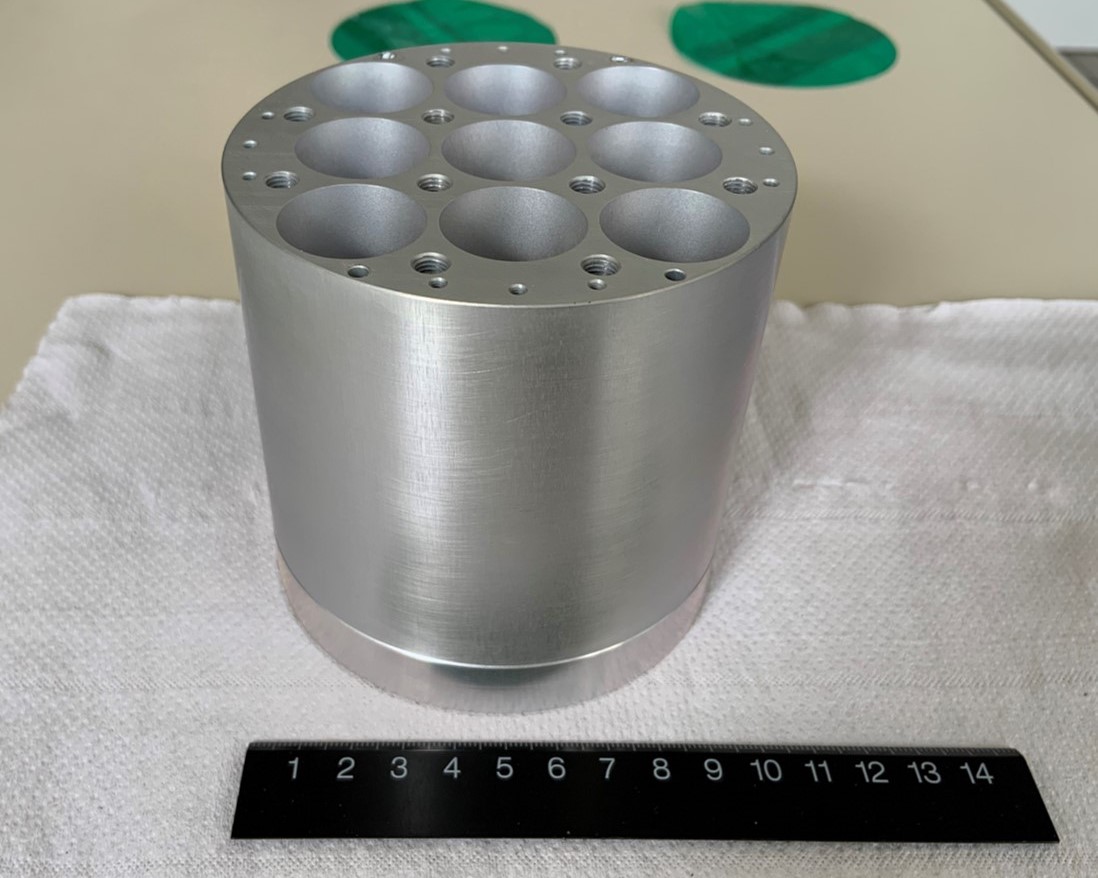}
    \includegraphics[height=4cm]{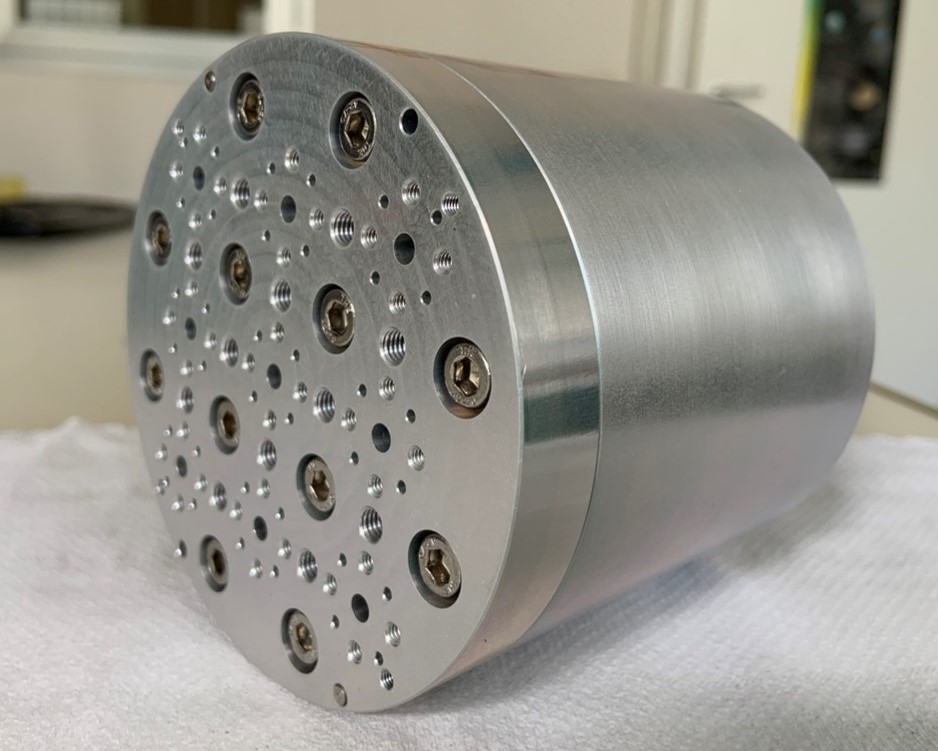}
   \end{tabular}
   \end{center}
   \caption[example] 
   {\label{fig:array2_pictures}Front, side, and back view of the COSMO 250 GHz array.}
   \end{figure}

The COSMO antenna system consists of two cylindrical arrays, shown in Fig.~\ref{fig:array1_pictures} and Fig.~\ref{fig:array2_pictures}, with a 10.1 cm diameter, each containing nine smooth-walled multi-mode feed-horns arranged on a square footprint with a 26 cm center-to-center distance. The low-frequency array operates in the $120-180$ GHz range, and the high-frequency one in the $210-300$ GHz range, to avoid the atmospheric emission lines and maximize the frequency coverage where the atmosphere is transparent to the microwave signal.

Whereas standard single-mode feed-horns terminate with a circular waveguide that selects only the fundamental TE$_{11}$ mode, multi-mode horns have a larger output circular waveguide that propagates several higher-order modes. The beam pattern, which describes the angular distribution of received/emitted radiation, results from the incoherent combination of all the individual modes. Since higher-order modes may have a non-Gaussian beam, they provide power further away from the antenna axis than in the single-mode case, thus generating a multi-mode beam pattern characterized by a broad and flat main beam, followed by a series of side lobes.

\subsection{The feed-horn design}
\label{subsec:design} 
\begin{figure} [ht]
   \begin{center}
   \begin{tabular}{c c} 
   \includegraphics[height=5cm]{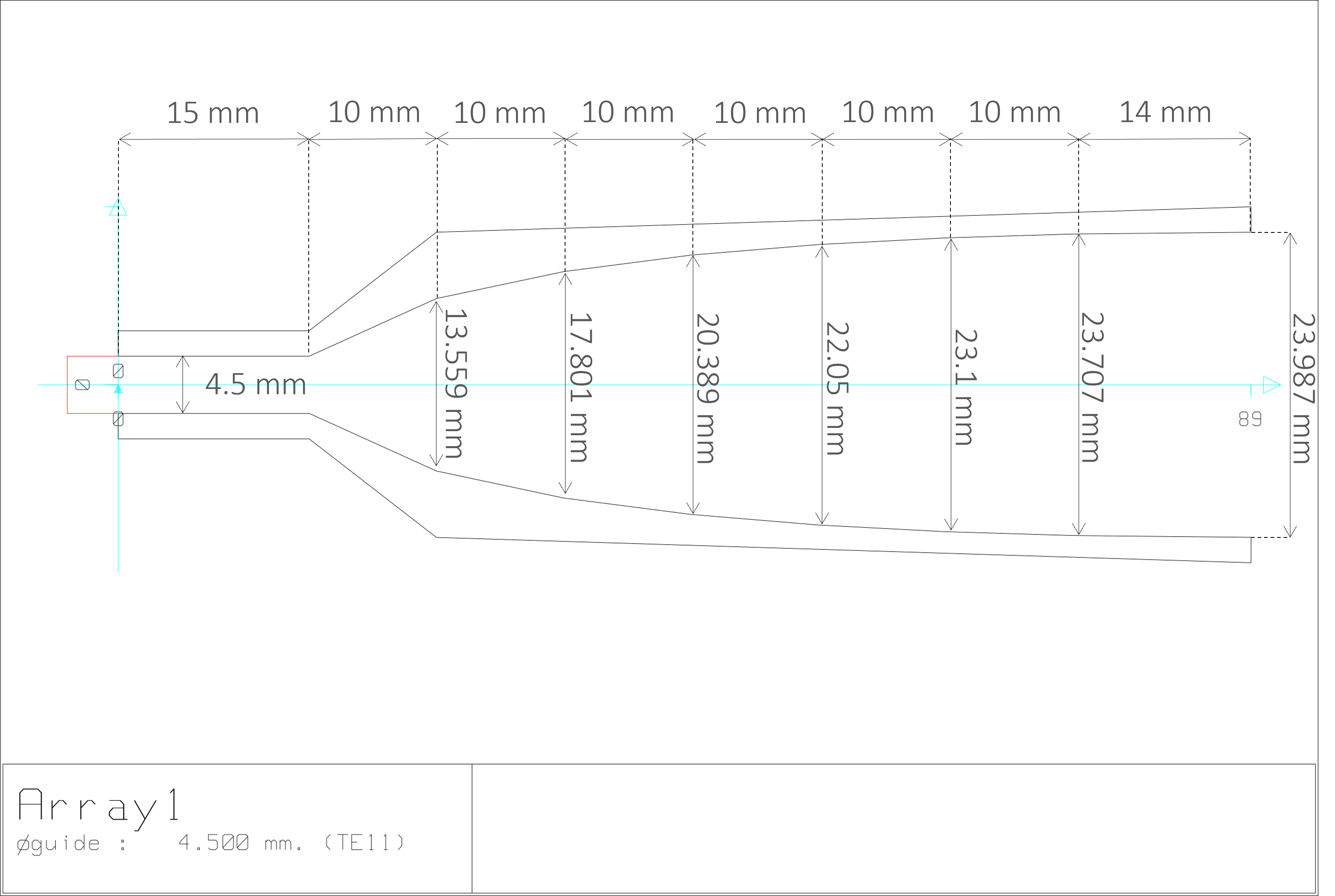}
    \includegraphics[height=5cm]{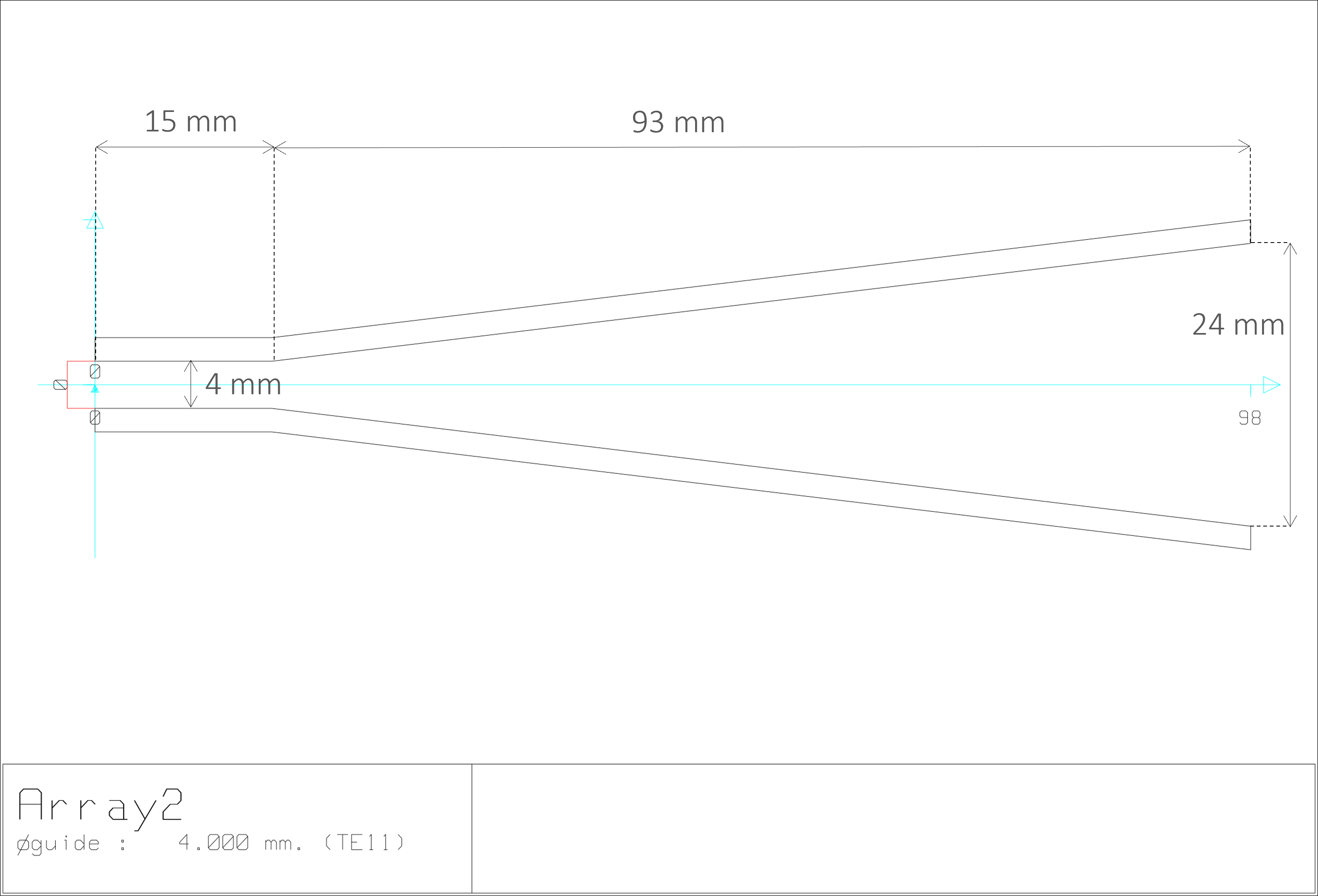}
   \end{tabular}
   \end{center}
   \caption[example] 
   {\label{fig:design}Profile of the COSMO feed-horns. \textit{Left panel:} the low-frequency horn is a Winston cone approximated by seven linear sections, terminated with a 4.5 mm waveguide, and is $89$ mm long. \textit{Right panel}: the high-frequency horn has a linear profile, terminated with a 4 mm waveguide, and is $108$ mm long.}
   \end{figure}

The low-frequency and high-frequency antenna profiles are shown in Fig.~\ref{fig:design}. The low-frequency horn is a Winston cone, while the high-frequency horn has a linear profile. Their design has been optimized to comply with mechanical and electromagnetic constraints, summarized in Table~\ref{tab:costraints}. The low-frequency horn terminates with a $4.5$ mm circular waveguide, to allow the propagation of up to 19 modes. The high-frequency one has a $4$ mm waveguide, for a maximum of 42 propagating modes.

We decided to manufacture the horns through CNC milling, a relatively fast and low-cost technique. We could manufacture a linear profiled horn as a single piece, and a more complex profile only if divided into up to 10 linear segments of at least $1$ cm thickness. For this reason, we approximated the Winston cone's paraboloid with seven linear steps. Since COSMO will perform only total intensity measurements, we chose a smooth-walled profile, which is easier to manufacture than a corrugated one.



The arrays are made in Ergal (Al7075), to conform with the rest of the focal plane, and each section of the horn profile is manufactured on a separate plate. The arrays are then obtained by superimposing the plates through dowel pins and tightening them with screws. Additional pin and screw holes are placed on the aperture plate to interface the array with a bandpass filter, and on the bottom plate to connect the flare flange. An interface to a standard UG387/U-mod flange is also accommodated around each circular waveguide to permit the beam pattern measurement.

   \begin{table}[ht]
    \caption{Summary of the mechanical and electromagnetic constraints driving the COSMO feed-horn design.} 
    \label{tab:costraints}
    \begin{center}       
    \begin{tabular}{|l|l|} 
    \hline
	\rule[-1ex]{0pt}{3.5ex} \textbf{Mechanical requirements} &  \\
    \hline
    \rule[-1ex]{0pt}{3.5ex}  Center-to-center distance & 26 mm  \\
    \hline
    \rule[-1ex]{0pt}{3.5ex}  Maximum aperture & 24 mm   \\
    \hline
    \rule[-1ex]{0pt}{3.5ex}  Maximum length & 20 cm  \\
    \hline
    \rule[-1ex]{0pt}{3.5ex}  Maximum number of linear segments & 10  \\
    \hline
    \rule[-1ex]{0pt}{3.5ex}  Minimum segment thickness & 1 cm  \\
    \hline 

	\rule[-1ex]{0pt}{3.5ex} \textbf{Electromagnetic requirements} &  \\
    \hline  
    \rule[-1ex]{0pt}{3.5ex}  Number of modes at $150$ GHz & $10-19$  \\
    \hline 
    \rule[-1ex]{0pt}{3.5ex}  Number of modes at $250$ GHz & $23-42$  \\
    \hline 
    \rule[-1ex]{0pt}{3.5ex}  FWHM & $\sim 17^{\circ}$  \\
    \hline     
    \end{tabular}
    \end{center}
    \end{table}

\subsection{The simulated multi-mode beam pattern}

We simulated and optimized the COSMO feed-horns with CST Microwave Studio, a commercial software suited for simulating multi-mode smooth-walled horns.  
We used the Integral solver, which is the fastest among all the available ones ($\sim 2$ min./mode$\cdot$frequency) because it solves the electromagnetic field equations within elementary cells covering only the antenna surface and not the entire metallic volume.

\begin{figure} [ht]
   \begin{center}
   \begin{tabular}{c} 
    \includegraphics[height=7cm]{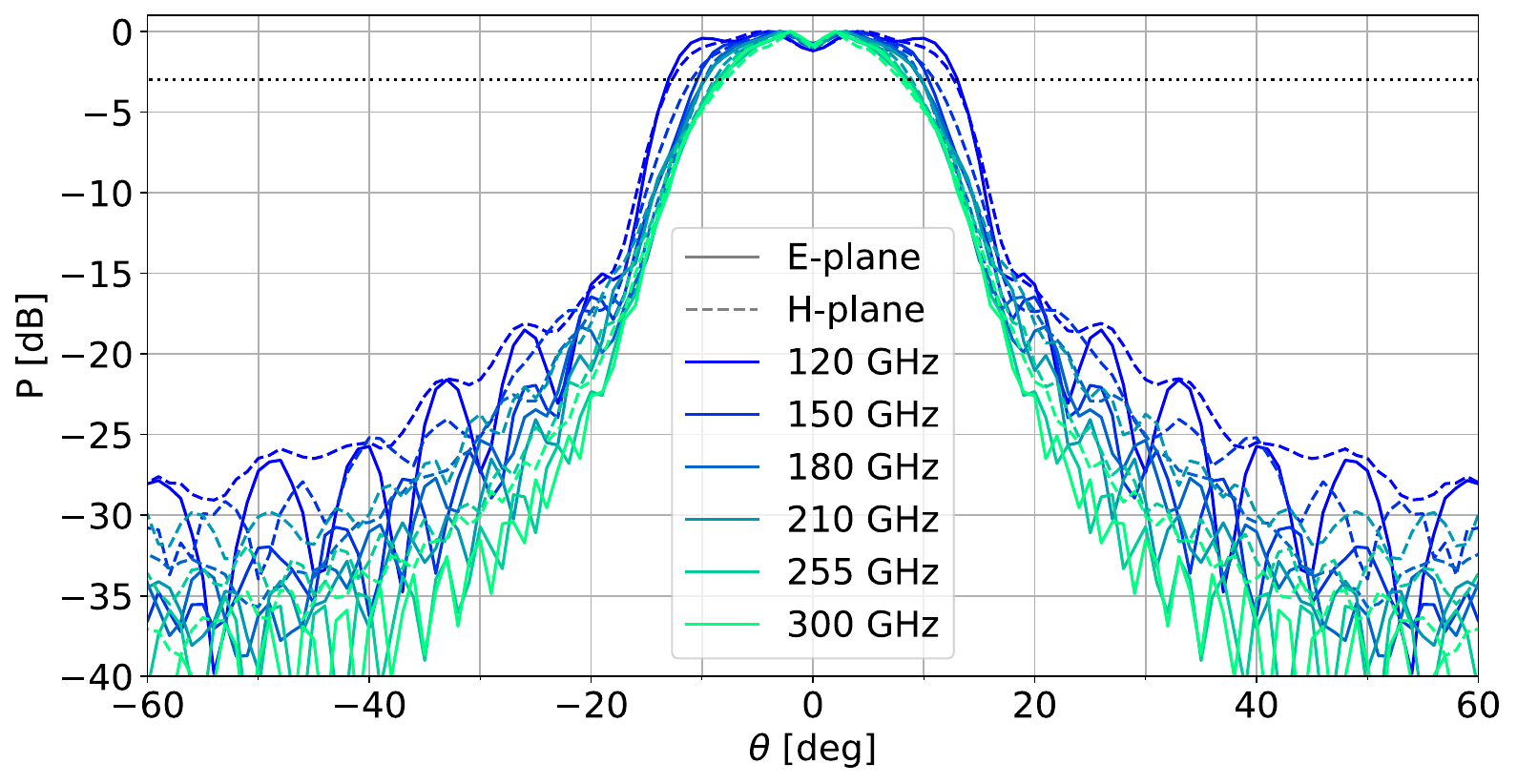}
   \end{tabular}
   \end{center}
   \caption[example] 
   {\label{fig:multimode_beam}Simulated multi-mode beam pattern for the two COSMO arrays. The E-planes (cuts at $\phi=90^{\circ}$) are the solid lines, whereas the H-planes (cuts at $\phi=0^{\circ}$) are the dashed lines. For each band, we show the beam pattern at the lower, center, and upper frequencies. The horizontal dotted line is the -3 dB level.}
   \end{figure}

The expected multi-mode beam pattern is obtained by incoherently summing up the power from all the propagating modes and is shown in Fig.~\ref{fig:multimode_beam}. The FWHM is around $20^{\circ} - 26^{\circ}$ for the low-frequency array and is around $16^{\circ}$ for the high-frequency one. The first side lobes are below -15 dB and the far side lobes are below -30 dB.

\section{Single-mode beam pattern measurements}
\label{sec:measurement} 

To test the antenna response, we performed beam pattern measurements in the Milano University anechoic chamber at room temperature using a system based on waveguide components and coherent propagation. The limiting factor of this procedure is that we can only measure the beam pattern of the fundamental mode because higher-order modes cannot propagate inside the receiving waveguide chain. Therefore, we consider this a first-level characterization of the COSMO horns. A comprehensive multi-mode characterization could be achieved at cryogenic temperature by coupling the feed-horns with incoherent detectors, and we defer it to future work. Broadband beam measurements will also be performed at the observing site using celestial sources.


\subsection{Experimental setup}
\label{subsec:method_and_setup} 

\begin{figure} [ht]
   \begin{center}
   \begin{tabular}{c c} 
   \includegraphics[height=5cm]{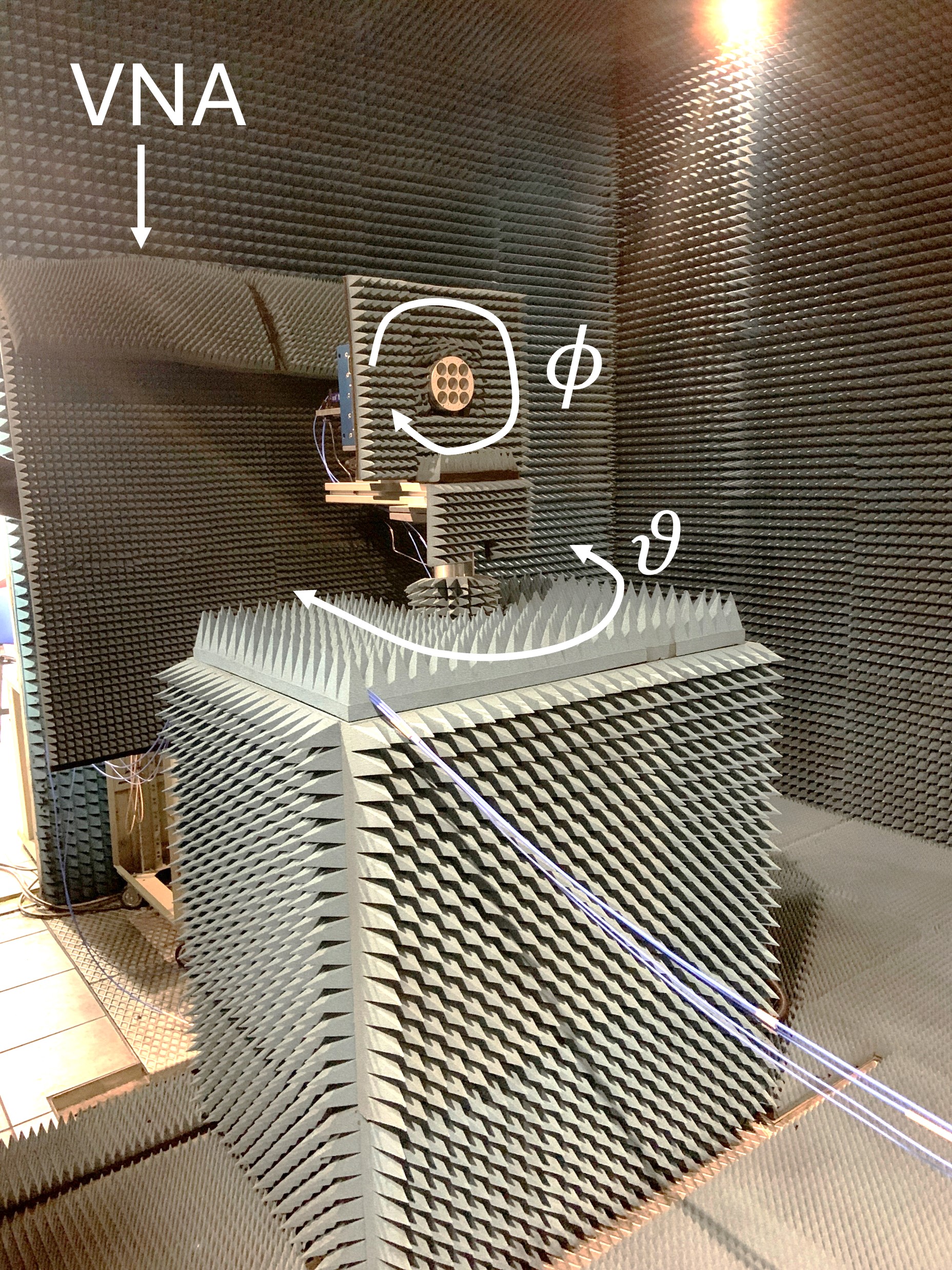} 
    \includegraphics[height=5cm]{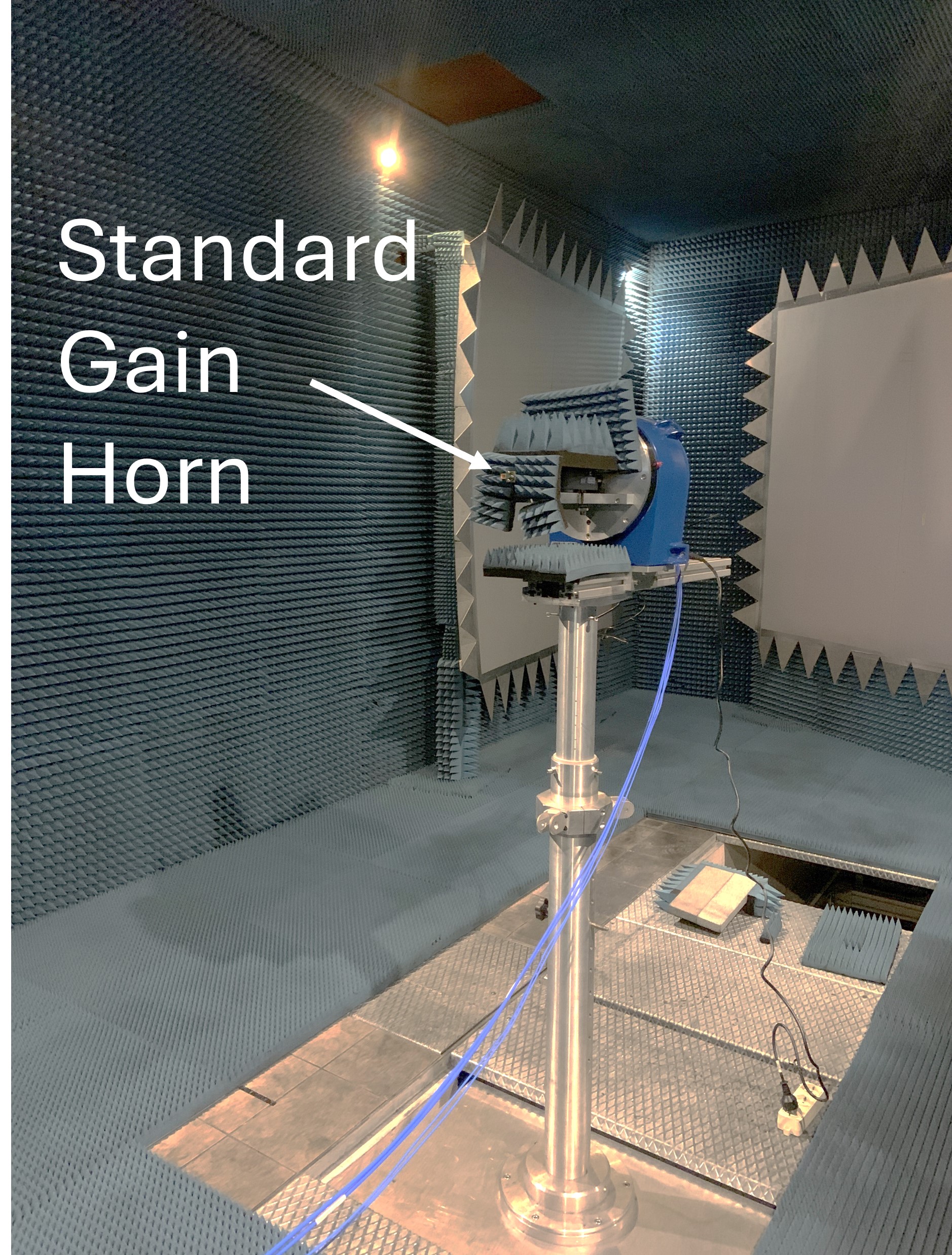}
   \end{tabular}
   \end{center}
   \caption[example] 
   {\label{fig:exp_setup}Far-field setup inside our anechoic chamber at the Physics Department, University of Milano. \textit{Left panel:} The COSMO array is mounted on a rotary table, covered in Eccosorb, that rotates over $\phi$ and performs azimuth scans. The VNA (not visible) is placed on an electronic rack behind, completely covered in Eccosorb. \textit{Right panel:} the source, a standard gain horn, is placed on a fixed pillar in front of the COSMO horn, at $\sim 2$ m distance, and covered in Eccosorb.}
   \end{figure}

\begin{figure} [ht]
   \begin{center}
   \begin{tabular}{c c c} 
  \includegraphics[height=5cm]{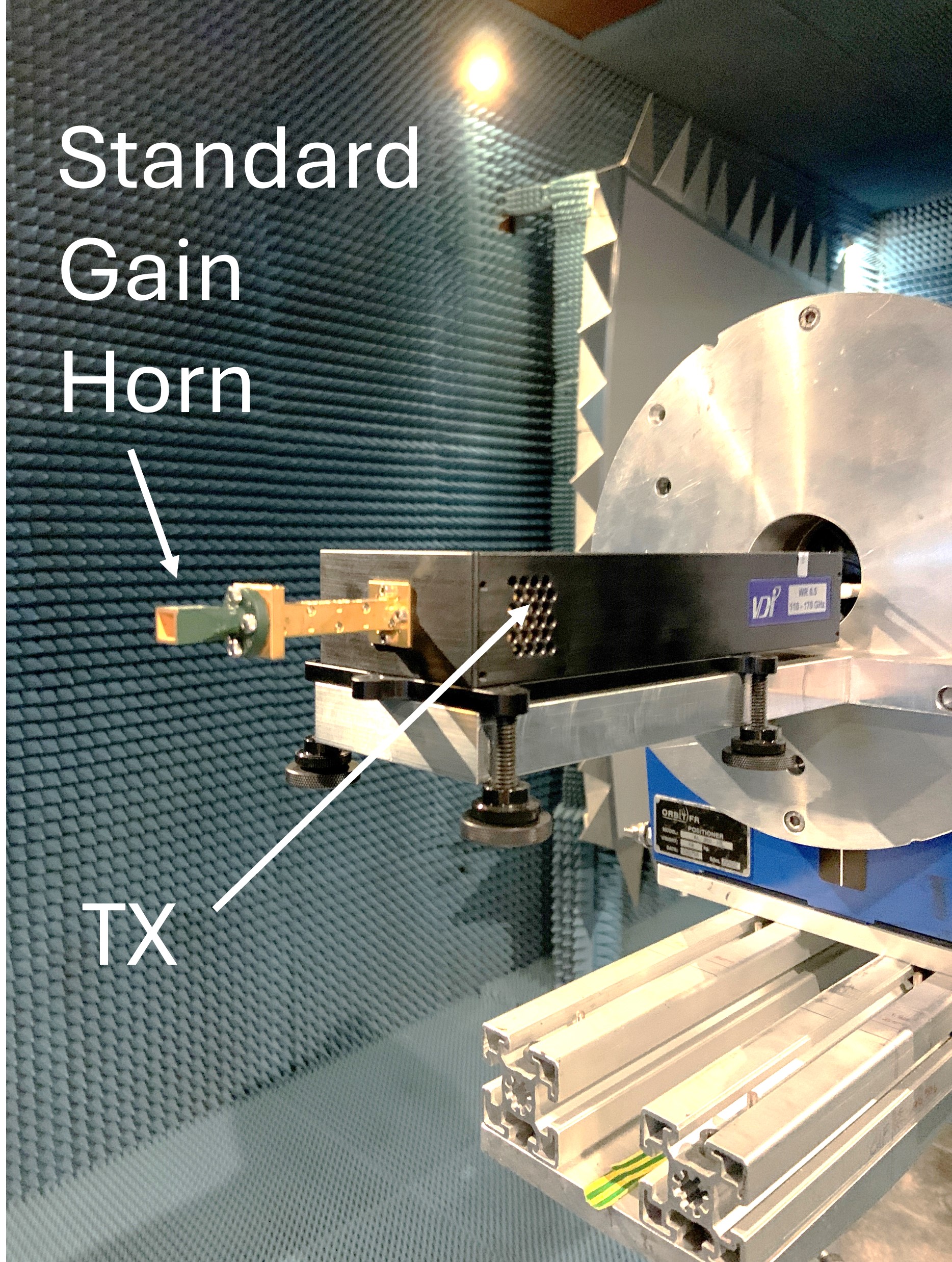} 
   \includegraphics[height=5cm]{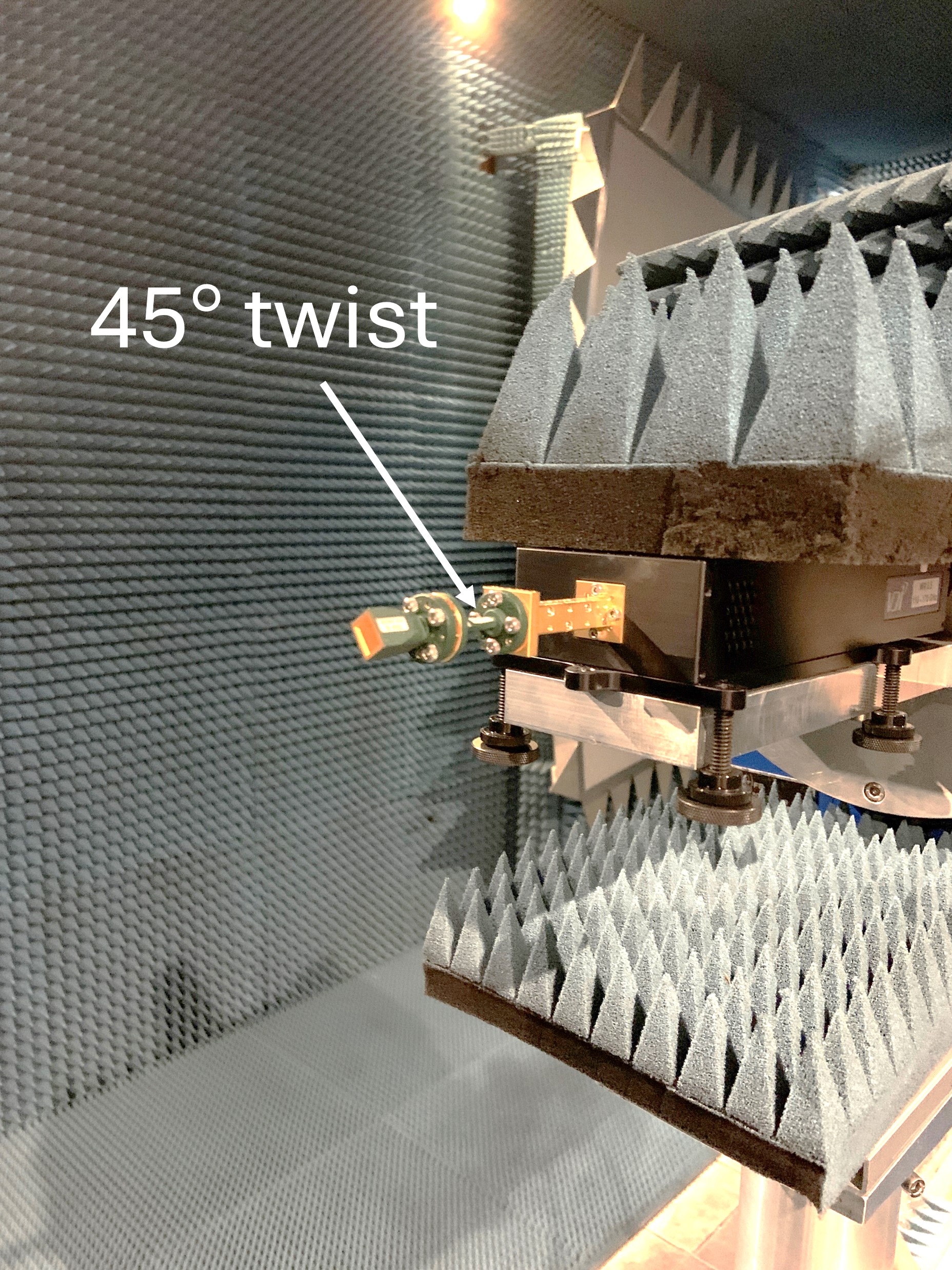}
  \includegraphics[height=5cm]{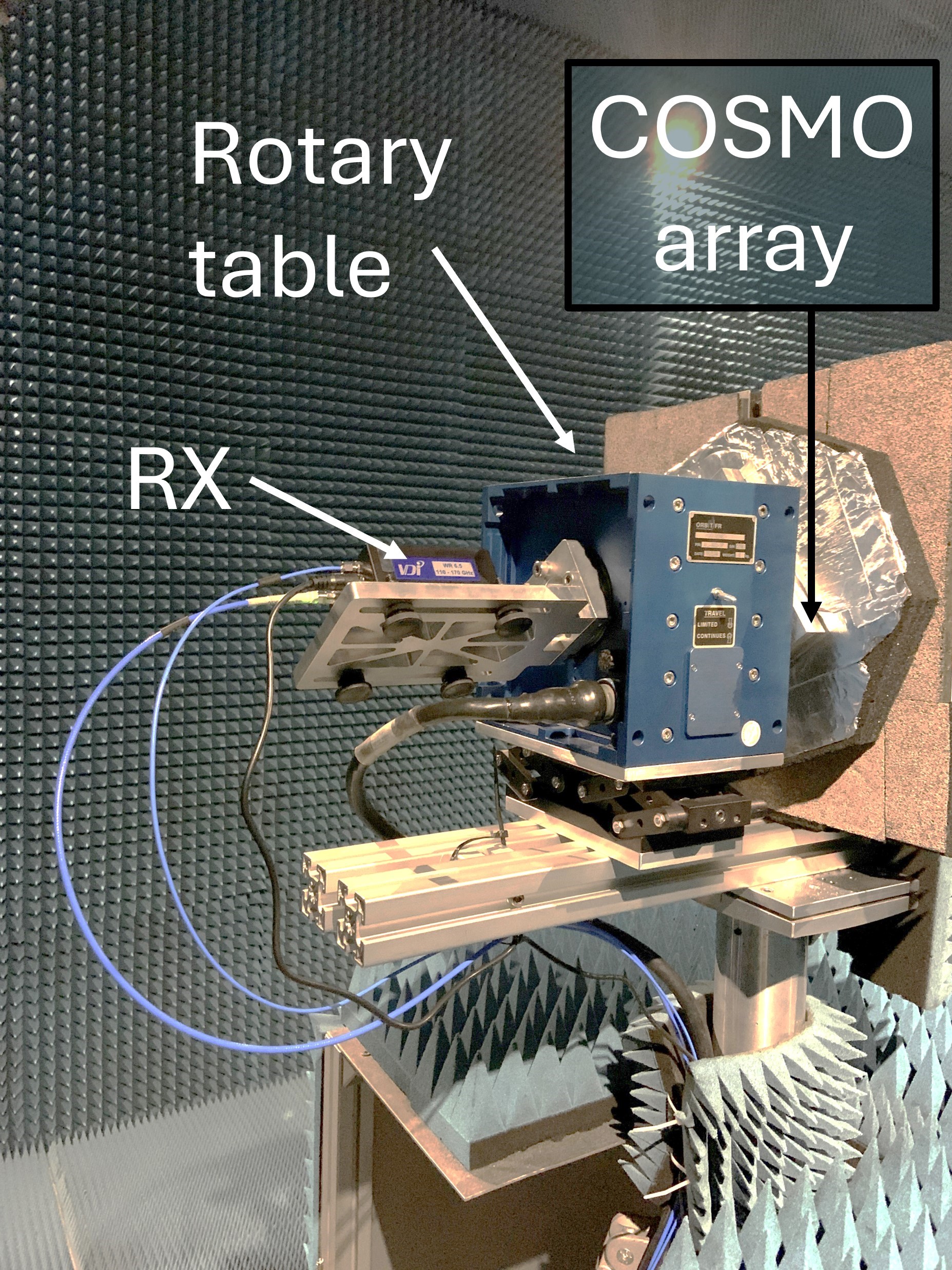}
   \end{tabular}
   \end{center}
   \caption[example] 
   {\label{fig:TX_RX_setup}Close up on the transmitting and receiving system. \textit{Left panel:} the pyramidal standard gain horn is connected to the TX frequency multiplier (in black). By default, the input polarization is vertical. \textit{Center panel:} a $45^{\circ}$ twist is interposed between the source and the TX to rotate the input polarization. \textit{Right panel:} the COSMO array (not visible) is mounted in front of the rotary table (in blue). A waveguide chain (not visible) propagates the signal inside the table and to the RX frequency multiplier. The RX is mounted on a built-in mechanical support at the back of the table.}
   \end{figure}

The experimental setup inside the anechoic chamber is shown in Fig.~\ref{fig:exp_setup} and~\ref{fig:TX_RX_setup}, and allows measuring one feed-horn at a time. The COSMO horn is mounted on a rotary table that rotates the system over $\phi$ and scans the beam pattern in azimuth with a user-defined angular step, $\Delta\theta$. The signal is transmitted by a pyramidal standard gain horn, placed at a far-field distance of $\sim 2$ m.

The source and the COSMO antenna are connected to a Vector Network Analyzer (VNA), from Agilent, through a frequency multiplier, from Virginia Diodes (VDI) operating in the $110-170$ GHz range. The transmitting frequency multiplier, TX, upconverts the low-frequency signal from the VNA to high-frequency towards the source, whereas the receiving one, RX, downconverts the output signal from the COSMO horn back to low-frequency towards the VNA. For each angular step of the scan, the VNA measures the amplitude and phase of the output signal.

The signal propagates in the TX and RX with a fixed polarization, selected by their rectangular waveguide. Therefore, we measured the beam pattern by rotating the receiver over $\phi$ until the two rectangular waveguides were aligned and scanning over $\theta$ at that given $\phi$ cut. 
Moreover, the default incoming polarization was $90^{\circ}$ (vertical), as shown in the left panel of Fig.~\ref{fig:TX_RX_setup}. Therefore, we used a $90^{\circ}$ twist and a  $45^{\circ}$ twist, interposed between the gain horn and the TX, to rotate the polarization, as shown in the center panel of Fig.~\ref{fig:TX_RX_setup}. This allowed us to measure the $0^{\circ}$ cut (horizontal), the $\phi = 45^{\circ}$, and also the $\phi = 135^{\circ}$ by using the two twists together. 

\begin{figure} [ht]
   \begin{center}
   \begin{tabular}{c} 
   \includegraphics[height=5cm]{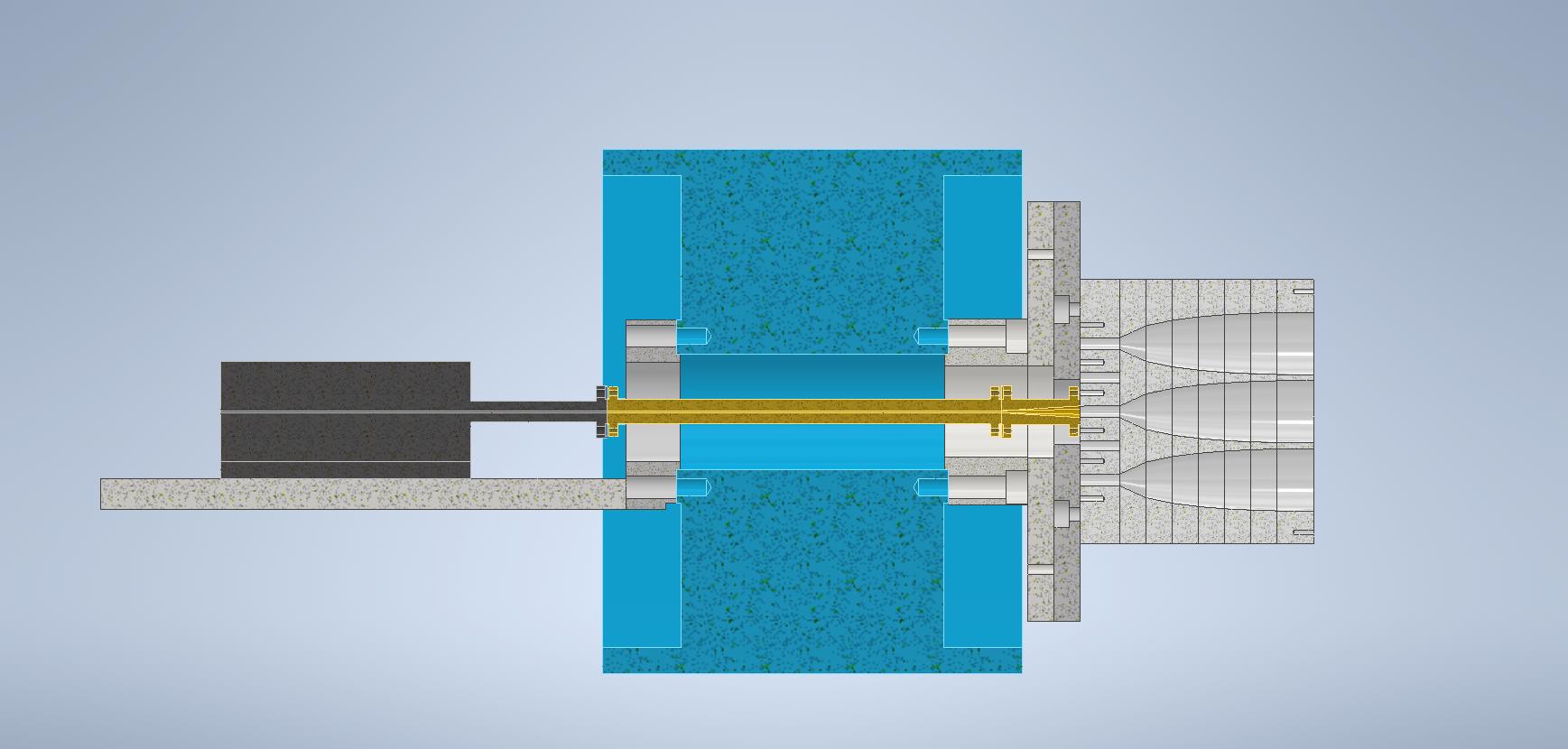}
   \end{tabular}
   \end{center}
   \caption[example] 
   {\label{fig:CAD_receiver}CAD semi-section view of the receiving chain. One
    antenna at a time is aligned with a rotary table (in blue) through a system of built-in mechanical interfaces. A waveguide chain (in gold), consisting of a custom circular multi-mode to rectangular transition and a rectangular waveguide straight, propagates the signal inside the table and to the RX frequency multiplier (in black). Higher modes are cut off at the rectangular waveguide.}
   \end{figure}

Since the RX receives the signal through a rectangular waveguide constructed to propagate only the fundamental mode, any other mode is reflected. Specifically, we used a custom circular multi-mode to rectangular transition, followed by a rectangular waveguide straight, to propagate the signal from the horn to the RX, as shown in the CAD semi-section view of Fig.~\ref{fig:CAD_receiver}. The mode suppression therefore occurs towards the end of the transition. 

\subsection{Preliminary results}
\label{subsec:results_discussion} 

\begin{figure} [ht]
   \begin{center}
   \begin{tabular}{c} 
  \includegraphics[height=16cm]{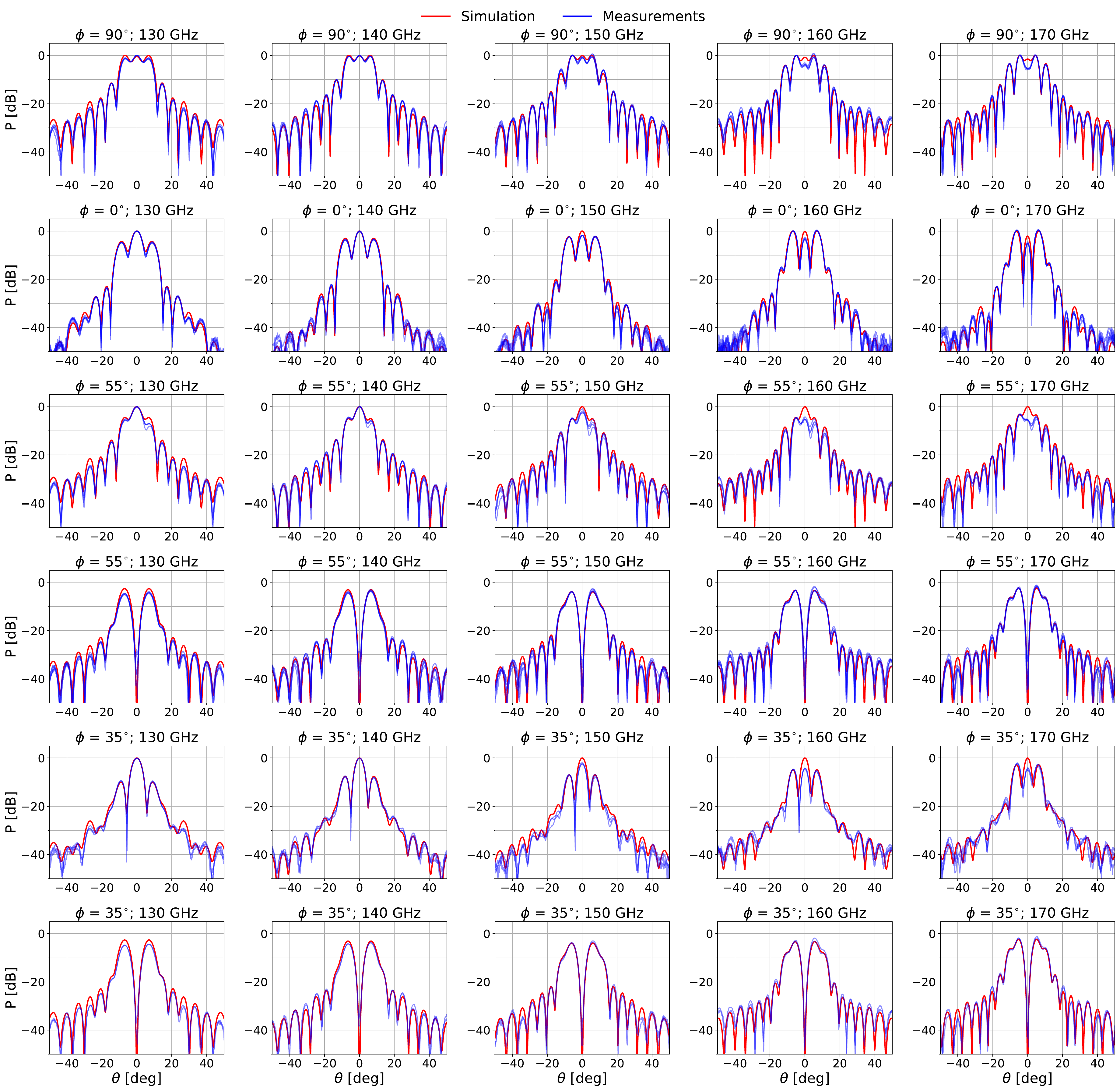}
   \end{tabular}
   \end{center}
   \caption[example] 
   {\label{fig:Meas_vs_Sim}Comparison between measurements in blue and simulation in red for the fundamental mode of five feed-horns in the 150 GHz array. Both the copolar and cross-polar beams are shown for the $55^{\circ}$ and $35^{\circ}$ cuts. Please note that we normalized the simulation to its maximum power, as is conventionally performed, but, to highlight the loss of directivity around the axis at higher frequencies, we re-scaled the measurements to align with the simulation on the first side lobe.}
   \end{figure}

We measured the beam pattern of the fundamental mode at $0^{\circ}$, $90^{\circ}$, $45^{\circ}$, and $135^{\circ}$ for five feed-horns in the 150 GHz array, namely: the central horn and four off-axis horns. For each cut, we measured the beam at five frequencies within the band: 130, 140, 150, 160, and 170 GHz. We found excellent repeatability and agreement with the simulations, as shown in Fig.~\ref{fig:Meas_vs_Sim}. 
We also identified a few non-idealities that we will further investigate, namely:
\begin{itemize}
    \item A $\sim 10^{\circ}$ net misalignment when measuring the $45^{\circ}$ and $135^{\circ}$ cuts, which resulted in the actual measurement of the $55^{\circ}$ and $145^{\circ}$, or equivalently $35^{\circ}$, cuts. This could be due to a defect in the $45^{\circ}$ twist, a horn-source misalignment, or a combination of both. We emphasize that during the calibration phase, we did not notice any major misalignment between the horn and the source, with or without the twist. According to our tests, the misalignment, if present, was within $1^{\circ}$. However, during the measurement campaign, an event occurred inside the chamber, which could have caused a mechanical push on the table, introducing the misalignment observed in the data.

    \item A $\sim 2-4$ dB loss of directivity around $\theta = 0^{\circ}$ above 150 GHz. We are still investigating this effect. Preliminary simulations indicate that a partial ($30\% - 40\%$) propagation of the first high-order mode within the receiving chain could be a possible explanation, and might be doable at higher frequencies because the cut-off frequency of the first higher mode in the rectangular waveguide is around 180 GHz. Another doable option is an imperfect source-horn coupling. 

    \item An asymmetry between positive and negative $\theta$ in the $55^{\circ}$ cut, which we do not observe in any other measured cuts and we, therefore, believe could be due to a defect in the measuring setup or an effect caused by the surrounding environment, but not from an actual defect of the horns. 
\end{itemize}

To investigate these hypotheses, we plan to repeat the measurements in a different anechoic chamber, completely covered in Eccosorb and with better control of the alignments. More in-depth simulations should also shed light on the cause of the directivity loss that we observe. Therefore, we are confident we can improve the already good agreement between measurement and simulation, and we will measure the 250 GHz array using the same procedure. 

\section{Conclusions}
\label{sec:conclusions} 

We described the design, development, and preliminary measurements of the $150$ GHz and $250$ GHz antenna arrays of the ground-based COSMO experiment, which aims at measuring the isotropic y-type spectral distortion of the CMB. 
Each antenna array consists of nine smooth-walled multi-mode feed-horns. The low-frequency horns have a step-linear profile and couple up to 19 modes onto a focal plane of KID detectors, while the high-frequency horns have a linear profile and propagate up to 42 modes. 
The two arrays are obtained by superimposing aluminum plates made with CNC milling. The simulated multi-mode beam pattern has a characteristic flat main beam with an FWHM ranging from $26^{\circ}$ to $16^{\circ}$. The side lobes are below -15 dB.

We performed beam pattern measurements in far-field conditions inside an anechoic chamber using a VNA. This strategy permits us to measure only the fundamental mode of the feed-horns since higher-order modes are cut off inside the receiving waveguide chain. We completed the measurements of the low-frequency array and found good repeatability and agreement with the simulations. 

We also identified three non-idealities to be further investigated, namely: 1. a $\simeq 10^{\circ}$ misalignment when measuring the $45^{\circ}$ and $135^{\circ}$ cuts, which we believe was the consequence of an unanticipated mechanical push that occurred during the measuring campaign; 2. a $2-4$ dB loss of directivity along the antenna axis above 150 GHz, which could be due to a partial propagation of the first higher-order mode or a defect in the source-horn coupling; 3. an asymmetry only in the $55^{\circ}$ cut, which we attribute to the measuring setup and not to an actual systematic effect of the horns.

We will repeat the measurements in a different anechoic chamber, completely shielded with Eccosorb and with better control of the alignments. Subsequently, the 250 GHz array will be measured using the same procedure. Given the high number of propagating modes, the multi-mode beam pattern measurement could be achieved at cryogenic temperature by coupling the feed-horns with incoherent detectors, and we defer it to future works. Once COSMO is operational, broadband beam measurements will also be performed at the observing site using celestial sources.

\acknowledgments 
This work has been funded by the PNRA (Programma Nazionale di Ricerca in Antartide) and the Italian Ministry of University and Research within the PRIN framework (Progetti di Rilevante Interesse Nazionale). The COSMO antenna arrays have been manufactured at the Pasquali SRL Workshop, in Milan.

\bibliography{biblio} 

\begin{thebibliography}{1}

\bibitem{Mater1990}
{Mather}, J.~C. et~al., ``{A Preliminary Measurement of the Cosmic Microwave
  Background Spectrum by the Cosmic Background Explorer (COBE) Satellite},''
  {\em The Astrophysical Journal Letters}~{\bf 354},  L37 (May 1990).

\bibitem{Planck2020}
Aghanim, N. et~al., ``Planck 2018 results: Vi. cosmological parameters,'' {\em
  Astronomy \& Astrophysics}~{\bf 641},  A6 (Sept. 2020).

\bibitem{Chluba_2011}
Chluba, J. and Sunyaev, R.~A., ``{The evolution of CMB spectral distortions in
  the early Universe},'' {\em Monthly Notices of the Royal Astronomical
  Society}~{\bf 419},  1294--1314 (12 2011).

\bibitem{Hill_2015}
Hill, J.~C. et~al., ``Taking the universe’s temperature with spectral
  distortions of the cosmic microwave background,'' {\em Physical Review
  Letters}~{\bf 115} (Dec. 2015).

\bibitem{Fixsen_1996}
Fixsen, D.~J. et~al., ``The cosmic microwave background spectrum from the
  fullcobefiras data set,'' {\em The Astrophysical Journal}~{\bf 473},
  576–587 (Dec. 1996).

\bibitem{Gervasi_2008}
Gervasi, M. et~al., ``Tris. ii. search for cmb spectral distortions at 0.60,
  0.82, and 2.5 ghz,'' {\em The Astrophysical Journal}~{\bf 688},  24–31
  (Nov. 2008).

\bibitem{Masi_2021_cosmo}
Masi, S. et~al., ``{The COSmic Monopole Observer (COSMO)},'' in [{\em {16th
  Marcel Grossmann Meeting on~Recent Developments in Theoretical and
  Experimental General Relativity, Astrophysics and Relativistic Field
  Theories}}{\nolinebreak\hspace{0.1em}]},  (10 2021).

\bibitem{Mele_2022}
Mele, L. et~al., ``Measuring cmb spectral distortions from antarctica with
  cosmo: Blackbody calibrator design and performance forecast,'' {\em Journal
  of Low Temperature Physics}~{\bf 209},  912--918 (Dec. 2022).

\end{thebibliography}
\bibliographystyle{spiebib} 

\end{document}